\documentclass[nonacm,sigconf]{acmart}

\setcopyright{none}
\renewcommand\footnotetextcopyrightpermission[1]{}

\definecolor{darkgreen}{rgb}{0.0, 0.65, 0.0}

\usepackage{listing}
\usepackage[]{algorithm2e}
\usepackage{graphicx}

\begin{document}
\date{}
\title[]{Zero Trust Real-Time Lightweight Access Control Protocol for Mobile Cloud-Based IoT Sensors}
\author{
{Atefeh Mohseni-Ejiyeh \\ 
University of California, Santa Barbara \\
Santa Barbara, California, USA \\
atefeh@ucsb.edu}
}

\maketitle
\pagestyle{plain}

\section*{Abstract}
\label{sec:abstract}
In IoT, smart sensors enable data collection, real-time monitoring, decision-making, and automation, but their proliferation exposes them to cybersecurity threats. Zero Trust Architecture enhances IoT security by challenging conventional trust models and emphasizing continuous trust verification in the overall \$875.0 billion IoT market projected by 2025.

This paper presents a new zero-trust real-time lightweight access control protocol for Cloud-centric dynamic IoT sensor networks. This protocol empowers data owners, referred to as sensor coordinators, to define intricate access policies, blending recipient identifiers and data-related attributes for data encryption. Additionally, the protocol incorporates efficient cryptographic primitives, eliminating the need for reliance on a trusted party. Furthermore, it ensures real-time data access while preserving data confidentiality and user privacy through seamless data upload to the cloud and the offloading of computationally intensive tasks from resource-constrained data owners and sensors. The protocol utilizes Merkle Trees for lightweight, ongoing trust measurement of sensors, ensuring efficient trust assessment by sensor coordinators. Simultaneously, the cloud conducts thorough trust evaluations for network entities including users. Comprehensive security analysis and performance evaluation highlight the protocol's effectiveness in tackling the multifaceted security challenges of IoT ecosystems while ensuring scalability and high availability.
\begin{sloppypar}
\section{Introduction}
\label{sec:intro}
Smart sensors, ranging from accessories to implants, play a crucial role in the IoT ecosystem, connecting to the Internet, collecting data, and aiding in intelligent decision-making. They seamlessly interface with smartphones and other devices, facilitating data collection and communication on the go. Leveraging advancements in low-power networks, electronics miniaturization, and sensor technology, these devices enhance efficiency, quality of life, and productivity. However, their increasing prevalence brings security challenges, as they become enticing targets for cybercriminals in both personal and industrial contexts \cite{wearables},\cite{iot2}.

The National Institute of Standards and Technology (NIST) has introduced a Zero Trust Architecture (ZTA), which serves as an enterprise's comprehensive cybersecurity strategy, incorporating essential elements such as workflow planning, component relationships, and access policies \cite{nist}. The intrinsic characteristics of IoT, including its heterogeneity, distributed nature, large-scale deployment, autonomy, dynamic behaviors, and safety considerations, position it as a highly favorable candidate for the application of these ZT principles \cite{blusky}.

In this context, the zero trust principle in IoT security takes center stage, emphasizing that trust should not be assumed based solely on physical or network location. IoT ecosystems, comprising billions of devices scattered across diverse locations and seamlessly integrated with cloud services, are inherently vulnerable to security threats due to their enlarged attack surface. Consequently, it becomes imperative to establish robust security protocols that transcend conventional trust paradigms, thereby fortifying the protection of these intricate networks.

In fact, three of the top 10 OWASP IoT vulnerabilities, including insecure data transfer and storage, lack of device management, and lack of physical hardening, could be effectively mitigated through the implementation of a robust and secure access control mechanism \cite{owasp}. However, despite the pressing need for such measures, there remains a noticeable absence of secure lightweight protocols designed specifically for IoT applications \cite{lack-light}. This underscores the critical importance of advancing research and development in this domain to address these vulnerabilities and fortify the security of IoT ecosystems through the implementation of a ZT lightweight access control protocol tailored for IoT sensors.

\subsection{Motivation}
The dynamic and security demands of IoT sensor networks necessitate an agile, robust data access control protocol. In these ever-changing, often challenging environments, real-time decision-making and data protection are paramount. To meet these requirements, a lightweight and flexible access control protocol is essential, ensuring efficient resource utilization and adaptability to evolving network conditions. Implementing a Zero-Trust model adds further complexity, demanding continuous monitoring, dynamic policy enforcement, and trust verification without prior assumptions. Our paper introduces a novel Zero-Trust Real-Time Lightweight Data Access Control Protocol for Cloud-centric IoT Sensor Networks, addressing these critical challenges.
\subsection{Our Contributions}
Our protocol introduces the following novelties:
\begin{itemize}
    \item \textbf{Zero-Trust Access Control}: Our protocol establishes a zero-trust environment, where trustworthiness of entities is verified on a per-request and per-resource basis. This entails continuous monitoring and secure storage of entities trust level.
    \item \textbf{Fine-Grained Access Control}: We empower data owners with the capability to define detailed access policies via fine-grained attribute-based encryption.
    \item \textbf{Lightweight Design}: Our protocol utilizes computationally efficient cryptographic primitives and eliminates the need for reliance on a trusted third party.
    \item \textbf{High Availability}: Our system ensures real-time data access by efficiently uploading data to the cloud and outsourcing computationally intensive tasks from resource-limited data owners to the cloud without compromising data confidentiality or users' privacy.
\end{itemize}
\subsection{Overview of Proposed Scheme}
In this section, we outline how our protocol achieves real-time, zero-trust access control for mobile IoT sensors.
\subsubsection{Zero-Trust Access}
We introduce a lightweight mechanism for sensor coordinators to manage sensors' trust scores. During initialization, sensors receive a full score based on authentication, protocol engagement, and user reports. A unique Merkle Tree is constructed, with the root hash shared as a seed trust token with the sensor. Trust scores are continuously updated during data transfer; sensors falling below a threshold lose access. We also employ dynamic trust for sensor coordinators, and users through a Trust-Level Evaluation Engine \cite{tlee} deployed in cloud, ensuring alignment with NIST Zero-Trust principles \cite{nist}.
\subsubsection{Fine Grained Access}
Our protocol combines Identity-Based Broadcast Encryption with Key-Policy Attribute-Based Encryption to provide data owners, known as sensor coordinators, with enhanced control and reduced operational burden through collaboration with cloud service providers. In our approach, sensor coordinators define the access policy during data encryption. This policy combines the recipient's identifiers (e.g., roles, geo) with specific data-related attributes (e.g., vital, urgent) using the "AND" operation. This design serves a dual purpose: it safeguards user identities from the cloud while preventing the cloud from decrypting the data. The cloud's role is limited to key generation for attributes and does not involve identity-related decryption keys.
\section{Design Overview}
\label{sec:solution}
In this section, we introduce the system model and outline the security prerequisites for public safety services, serving as a case study.

\subsection{Architecture}
\begin{figure}[!t]
  \centering
\includegraphics[width=\linewidth]{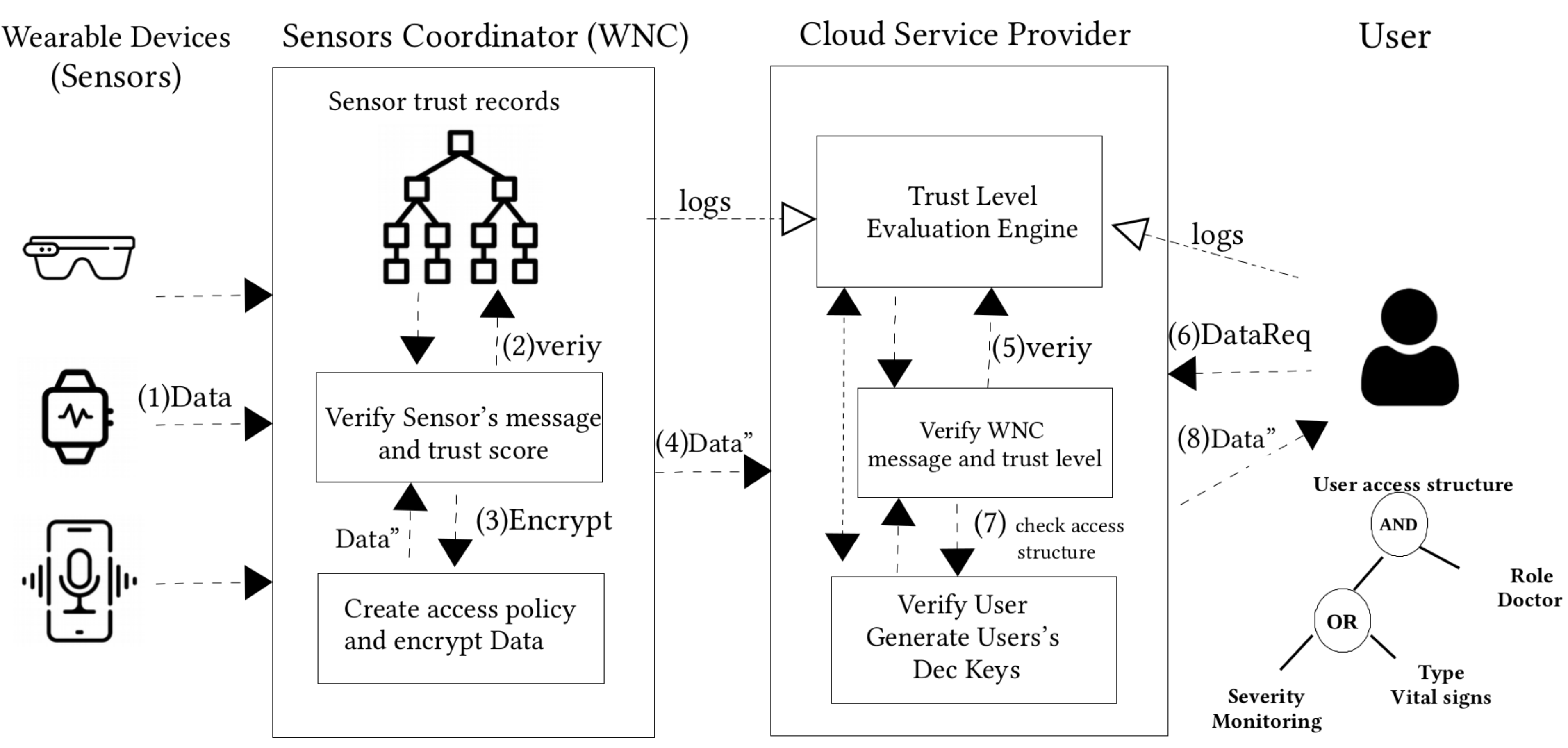}
\caption{Architecture of the proposed scheme} 
\label{fig:sysm}
\end{figure}

The principal system roles integral to the proposed protocol encompass the following entities as depicted in Figure \ref{fig:sysm}.

\begin{itemize}
\item Wearable Devices ($W_i$): These sensors are affixed to first responders' bodies, tasked with gathering vital information.

\item Wearable Network Coordinator: WNC assumes the role of managing the ($W_i$s) and collecting their data. 

\item Cloud Service Provider (CSP): Operating as an access control authority mediating between users and WNCs, the CSP is responsible for authenticating users seeking access to Wis data and trust level evaluations.

\item Users ($U_i$): These individuals, often Command and Control officers geographically distributed, oversee and collaborate with first responders in real-time.

\end{itemize}

\subsection{Trust model and threats}
In our approach, we adhere to the "never trust, always verify" principle, which requires rigorous verification of the identity, device integrity, and security posture of all network entities, including users and devices, before granting access to resources. Following the Dolev-Yao threat model \cite{dolev}, an adversary possesses the capability to alter or delete the message contents transmitted over the insecure public channel. The Cloud Service Provider operates on the "honest-but-curious" premise, authorizing users and generating private keys based on attributes while preserving the potential for data inspection. Users pose insider threats, potentially attempting unauthorized data access either independently or collaboratively. Security concerns also extend to WNC and $W_i$s due to potential compromise by adversaries, risking the introduction of falsified or malicious data into the cloud platform, endangering data integrity and authenticity.

\subsection{Design Goals}
Our objective is to design a scalable, real-time access control mechanism within the framework of zero trust. This protocol should effectively utilize device resources and accommodate dynamic network characteristics without compromising security. Here, we outline the main security requirements.

\subsubsection{\textbf{Zero Trust Access Control}}
Our aim is to incorporate a secure access control mechanism enabling data owners (WNCs) to share information securely with users via a cloud platform. This mechanism must empower data owners to enforce fine-grained access structures, ensuring data decryption aligns with specific attribute requirements. In a zero trust environment, trust is absent between entities, requiring WNCs to monitor $W_i$s behaviors, while CSP continuously assesses the trustworthiness of both WNCs and Users.

\subsubsection{\textbf{Mutual Authentication}}
A mutual authentication mechanism must be incorporated between $WNC$s and $W_i$s and also between WNCs and the CSP to fortify the system against impersonation attacks.

\subsubsection{\textbf{Data Confidentiality and Integrity}}
To safeguard against data leakage and message fabrication, the protocol must employ appropriate encryption and hash algorithms to ensure data confidentiality and integrity during data transmission and at rest among the involved entities.

\section{Proposed Zero-Trust Real-Time Access Control for IoT Sensors}
\label{sec:implementation} 
The proposed protocol comprises three phases as follows:

\subsection{Initialization}
This phase encompasses key agreement between $WNC$ and $W_i$s, as well as between $WNC$ and CSP. It enables $WNC$ to evaluate the trustworthiness of messages from $W_i$ and establish cryptographic schemes utilized in the data transfer phase. These cryptographic schemes facilitate $WNC$ in managing users' static attributes through Identity-Based Broadcast Encryption and handling users' dynamic attributes via Key-Policy Attribute-Based Encryption (KP-ABE).

\subsubsection{\textbf{Key Agreement}}\label{key-agg}
To establish key agreements between entities $WNC$ and $W_i$s, as well as between $WNC$ and $CSP$, we adopt an ECDH scheme to generate pairwise MAC keys, denoted as $k_{wnc,wi}$ and $k_{wnc,csp}$, following a similar approach as presented in \cite{ps-auth}.

At the conclusion of the key agreement handshake between each $W_i$s and $WNC$, a symmetric encryption key ($k$) is sent to each $W_i$s for generating a key hash chain \cite{h-chain} based on $k$, which will be used to encrypt data collected by $W_i$s.
Upon receiving $k$, each $W_i$ produces a key hash chain $\mathcal{H}={h_{n-1}, h_{n-2}, ..., h_1}$, where $h_0 = k$ and $h_i = H(h_{i-1}), 1<= i<=n$. Each $h_i$ is utilized to encrypt vital data recorded by $W_i$. Each $h_i$ has a validity period determined by $WNC$, matching the number of hash chain values for each epoch.

\subsubsection{\textbf{Seed Trust Token Distribution}}
The Wearable Network Coordinator employs an efficient real-time trust score management algorithm for $W_i$ sensors. During initialization, a seed token is assigned, computed using Eq. \ref{trust-score}. This equation combines authentication ($F_1$), activity ($F_2$), and user-reported event ($F_3$) factors, with respective weights assigned by scoring factors ($SF_i$): 40\% for $SF_1$ and $SF_2$, and 20\% for $SF_3$, prioritizing objective metrics.

\begin{equation}\label{trust-score}
TS = F_1 \times SF_1 + F_2 \times SF_2 + F_3 \times SF_3
\end{equation}

The WNC initializes with a score of 100 and constructs a Merkle Tree using the seed score and device ID as leaves, sharing the root ($TS_{i,0}$) with $W_i$. Subsequently, for any violated factors ($F_i$) by $W_i$, WNC reduces the $W_i$'s score until it falls below a threshold, with the maximum score set at 100.

\subsubsection{\textbf{Running KP-ABE Setup by WNC}}
\textbf{Step 1}: WNC selects a security parameter $k$ and an attribute set $U = U_1 \cup U_2 $ to initiate a KP-ABE setup. Here, $U_1$ corresponds to a receiver set ${S}$ of the IBBE scheme, while $U_2 $ is a set of chosen attributes by WNC for the KP-ABE scheme. The KP-ABE setup yields the public key set (PK) and master key set (MK).

WNC should then share the $PK$ and $MK=MK_1 \cup MK_2$ sets with CSP to enable the generation of users' private keys during the registration phase. It's important to note that CSP, being a semi-honest entity, must not have the ability to decrypt stored data. WNC transmits the $MK$ and $PK$ sets related to $U_2$ via $Enc_{(kwnc,csp)}(MK_2)$ and generates decryption keys for a receiver set $S$ and sends the encrypted version of the $SK$ through $Enc_{IBBE}(SK)$, using an IBBE scheme.

\textbf{Step 2}: Upon receiving a message from WNC, CSP first verifies the integrity of $h(*)$ and then proceeds to decrypt the $Enc_{(kw,c)}(MK_2)$ values. CSP maintains a mapping between user IDs ($U_1$) and their respective encrypted secret keys ($SK$).

\begin{figure}[!t]
  \centering
\includegraphics[width=1\columnwidth]{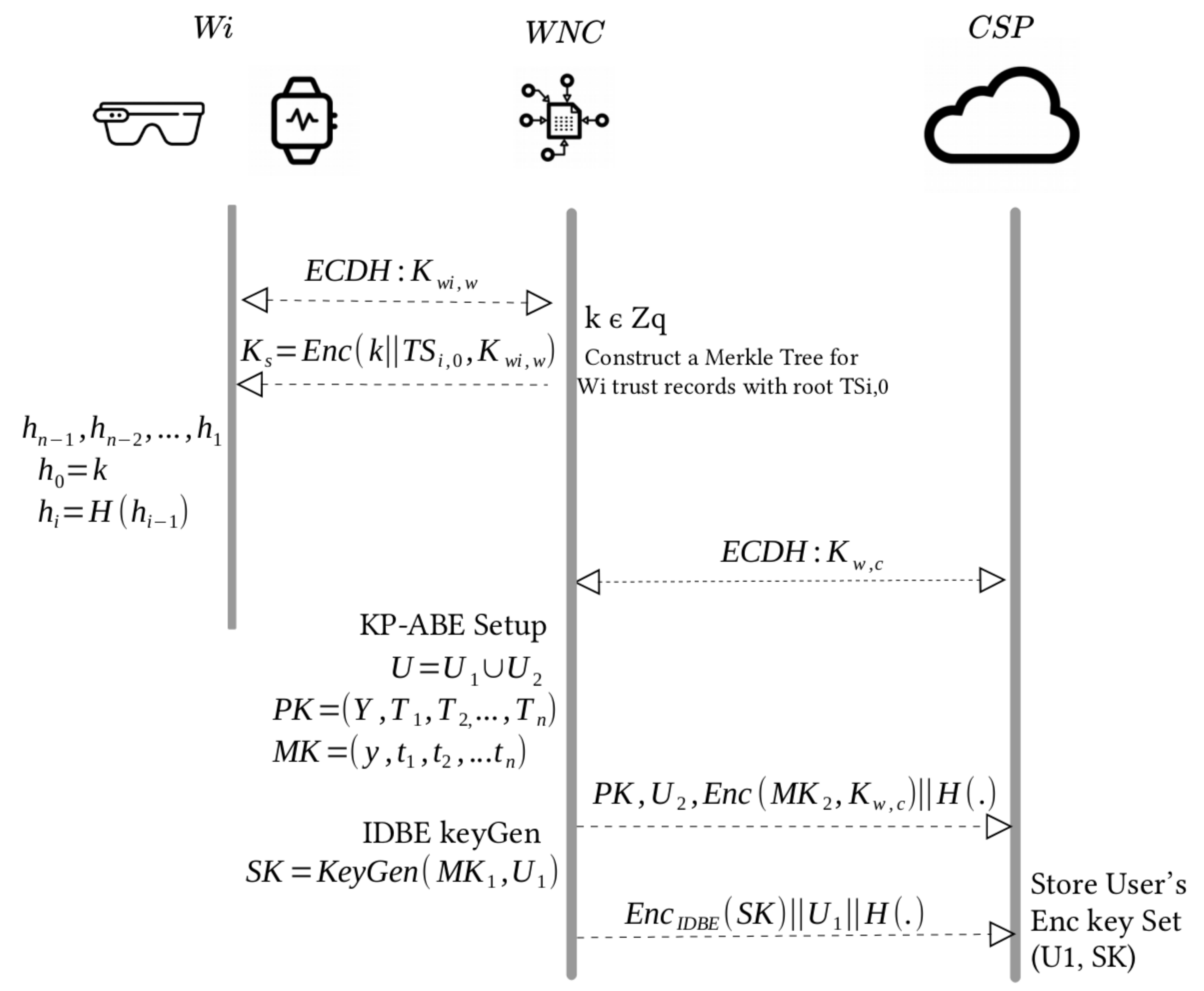}
\caption{Initialization phase} 
\label{fig:protp}
\end{figure}

\subsection{User Registration}
In this phase, users register by sending their IDs to the cloud, which associates private keys with their attributes.

\textbf{Step 1}: $User_i$ initiates the registration process by sending its ID to CSP, which is transmitted via $Enc_{pri-user}(ID)$. ID serves as an identifier for a group of users (e.g., Command and Control officers) to maintain privacy.

\textbf{Step 2}: CSP initiates the validation of the request, ensuring its authenticity. If the request is verified, CSP proceeds to execute a key generation algorithm using the user's predefined access structure P (maintained by CSP), the master key (MK), and the public key set (PK). Subsequently, CSP returns $Enc_{pub-user}(Sk_i)$ to the user, where $Ski$ encompasses the user's private key policies along with the {encrypted ID-based key $SK_{ID}$}. Additionally, CSP adds the user's ID to the User List, which maintains the roster of authorized users.

\textbf{Step 3}: $User_i$ receives the message, first verifying its authenticity, and then decrypts its private key policy. To access resources, the user combines their attribute-based subtree with the user ID subtree using a logical 'AND' operation. This integration results in a combined private key set that encompasses both the user's attributes and their user ID. Please refer to delegation of private keys section in \cite{kp-abe} for more details on access tree reconstructoin in KP-ABE protocol.

\begin{figure}[!t]
  \centering
\includegraphics[width=1\columnwidth]{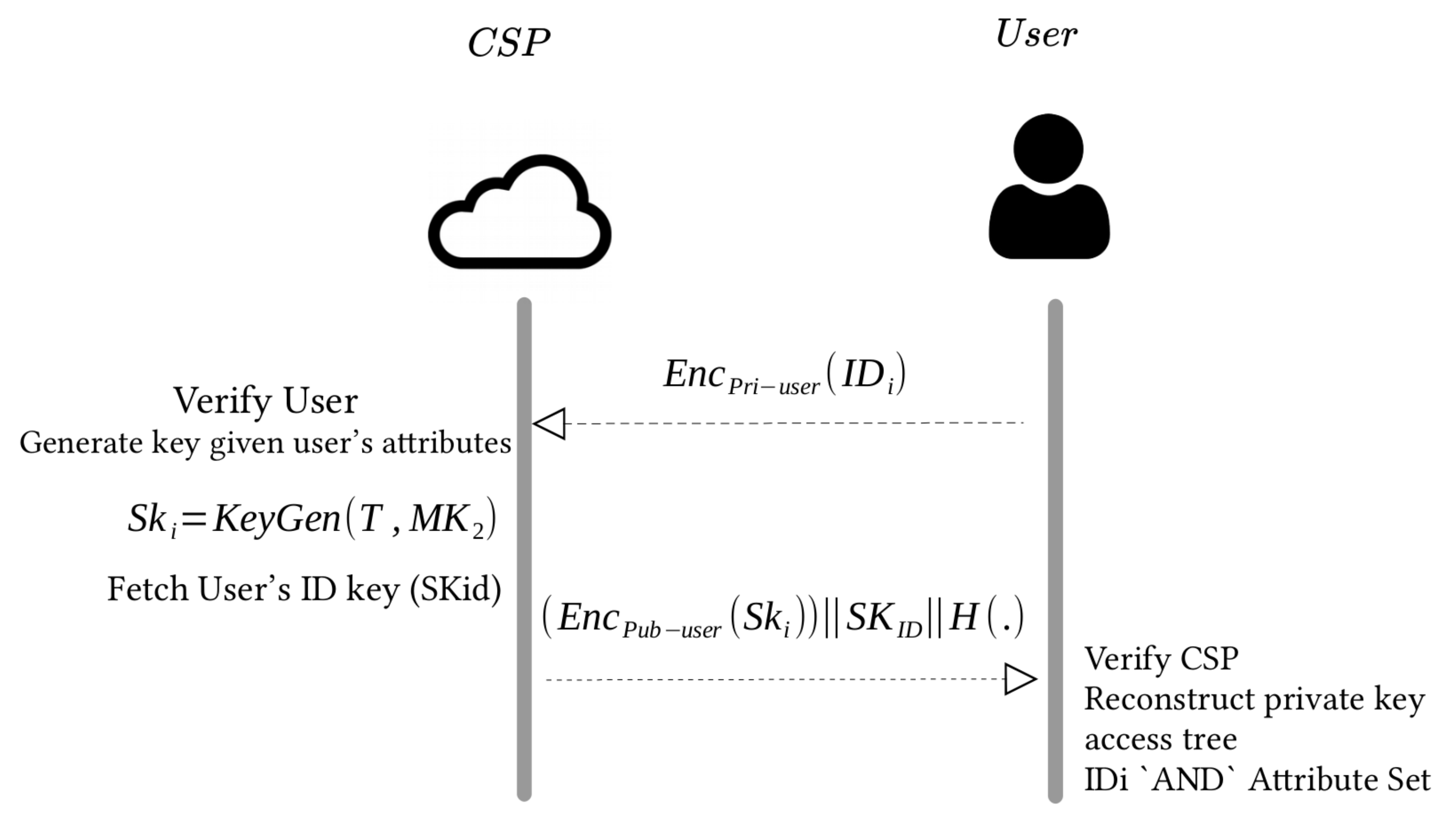}
\caption{User registration phase} 
\label{fig:protp}
\end{figure}
\subsection{Data Transfer}
This phase includes the periodic collection of encrypted data from $W_i$ sensors by the Wearable Network Coordinator (WNC), with the frequency depending on the specific needs of the public safety operation (ranging from seconds to minutes). The collected data is then encrypted using a KP-ABE encryption algorithm, and the encrypted data is subsequently uploaded to the cloud platform. Users are granted access to the data based on their authorized attributes, identifiers (e.g., roles), and trust levels. The following steps outline this phase.

\textbf{Step 1}: Each wearable sensor ($W_i$) encrypts its sensed data (M) using key $h_j$ from the key hash chain and sends $Enc_{h_j}(M)||TS_{i,j} || HMAC(*)$ to WNC.

\textbf{Step 2}: WNC verifies the HMAC value of the received message and awaits data from other $W_i$ sensors. This waiting period is typically about a few seconds or less, depending on the validity period of each key ($h_j$) in the key hash chain. WNC validates and if true updates the trust score's Merkle Tree for $W_i$ sensor and sends the root value to $W_i$ as $TS_{i,j+1}$.

\textbf{Step 3}: WNC selects a subset of attributes $I$ from the attribute set $U$ and executes the KP-ABE Encryption function with $I$ and $h_j$ (used as a message). WNC then transmits the encrypted file, as illustrated in Fig \ref{fig:protp}, along with the HMAC value of the message, using the shared key $Kwnc,csp$.

\textbf{Step 4}: The Cloud Service Provider (CSP) validates the data upload request from WNC, considering integrity and trust behavior aspects.

\textbf{Step 5}: When a $User_i$ intends to access the data, they send their certificate $Cert_i$ with desired attribute set $\lambda$  to the cloud.

\textbf{Step 6}:  CSP verifies the received certificate and checks if the user's ID exists in the User List. If the user's ID is found and their trust level is acceptable, access to the encrypted data is granted for messages matching $\lambda \in I$; otherwise, the user's request is disregarded.

\begin{figure}[!h]
\centering
\includegraphics[width=1.1\columnwidth]{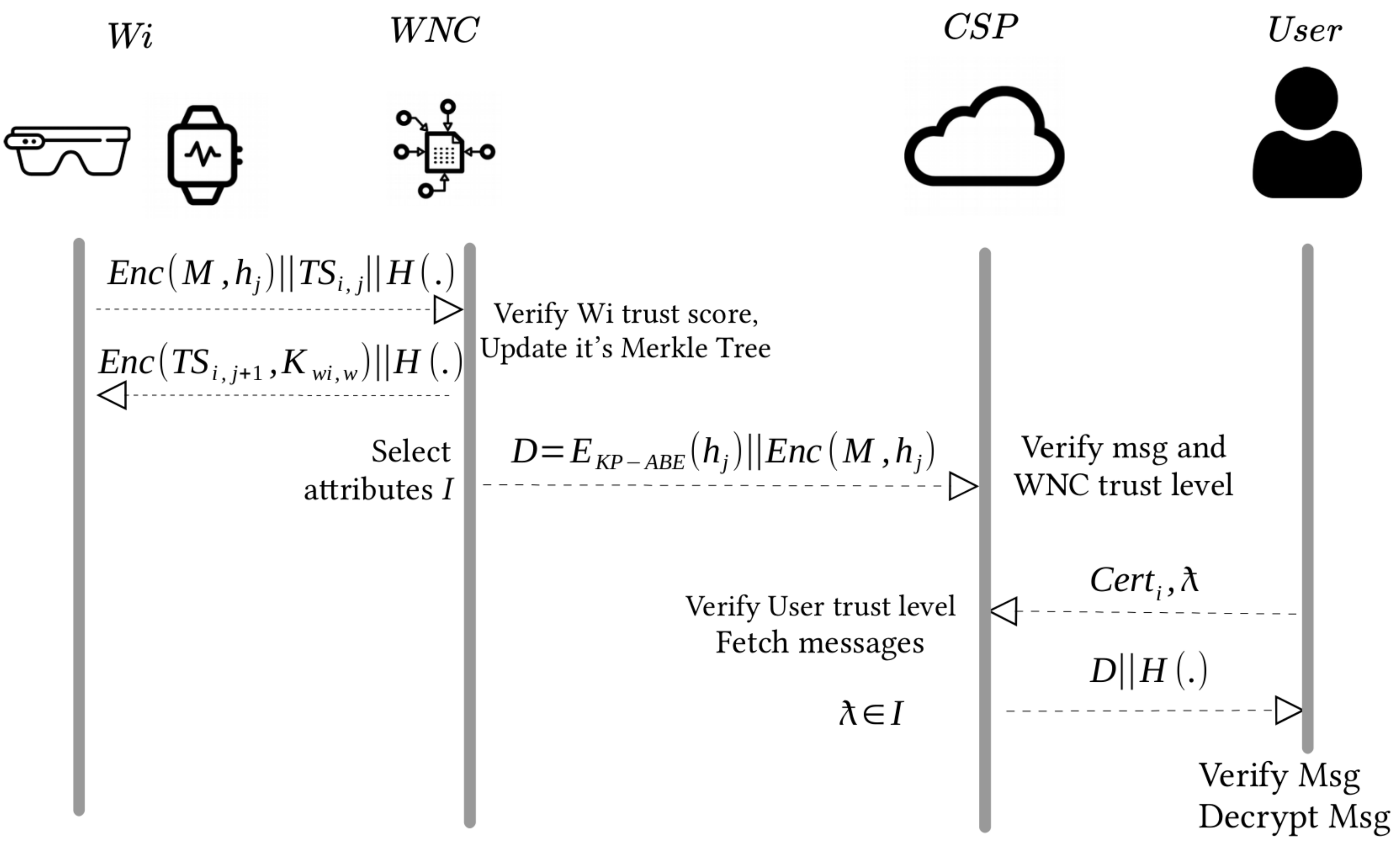}
\caption{Data transfer phase} 
\label{fig:protp}
\end{figure}
\section{Evaluation}
\label{sec:evaluation}
In this section, we thoroughly evaluate our proposed scheme, analyzing both computation costs and security aspects.

\subsection{Computation Overhead - Initialization and User Registration Phases}
Here, we provide an overview of the key computational overhead, highlighting that the overall computation cost is within acceptable limits.

\textit{ECDH Key Exchange}: During this phase, wearable nodes and WNC, as well as WNC and CSP, perform an ECDH scheme to compute a shared key.

\textit{Trust Score Tree Construction}: The WNC encrypts the seed of the key hash chain $K$ using $K_{w,w}$ and simultaneously constructs a Merkle Tree for $W_i$ trust scores, which includes the hash of the seed trust score and the sensor's ID as leaves.

\textit{Key Chain Generation}: Each wearable node must decrypt the received message $h_0 = K$ using the key $K_{w,w}$, and then generate a hash chain. The cost for this operation is $N * T_{SHA}$, where $N$ represents the length of the key hash chain. The length of the key hash chain is determined by WNC based on the operational requirements.

\textit{KP-ABE Setup}: WNC runs a KP-ABE setup with a chosen attribute set $U$ and security parameter $k$. This operation can be performed offline and does not introduce additional runtime overhead.

\textit{Share Public and Master Key Sets}: WNC uploads the generated public and master key sets ($PK$ and $MK$) to the cloud platform. WNC encrypts the master key set related to the chosen attribute set $U_1$ with its shared key $K_{w,c}$. Additionally, WNC generates decryption keys from the MK set related to $U_2$ with the corresponding IBBE key. The total computational cost for this step is represented as $T_{KP-Setup} + n * T_{IBBE-KeyGen}$,  where $n$ refers to the cardinality of the receiver set of the IBBE scheme.

\textit{User registration request}: The cloud generates user's private keys ($Sk_i$) at a cost of $N_1 * T_{KP-KeyGen}$, where $N_1$ represents the cardinality of the attribute set $U_1$. Additionally, there is an associated cost for public key encryption by the cloud to securely transfer the user's keys.

\subsection{Computation Overhead of Data Transmission Phase}
The computation overhead during the data transmission phase primarily involves encryption operations carried out by wearable nodes (Wis) and WNC. Specifically:

\textit{Wearable Nodes (Wis)}: Each wearable node computes a symmetric encryption over the sensed data, which incurs a computational cost denoted as $T_{Enc}$.

\textit{WNC}: WNC performs KP-ABE encryption on received packets with a computational cost denoted as $T_{KP-Enc}$. Verifying the trust of a sensor requires a single hash operation on the sensor's Merkle Tree.

\textit{CSP}: The cloud authorizes user data access requests by verifying whether the user's attributes satisfy the access policy on the decrypted data, alongside assessing the user's dynamic trustworthiness. This trust assessment is conducted by the Trust Level Evaluation engine \cite{tlee}.

The overall computation costs presented in Table \ref{tab:comm-cost}. The detailed trust evaluation by CSP is not quantified, as it imposes negligible overhead given the available cloud resources. Definitions for terms used can be found in Table \ref{tab:comm-terms}.

\begin{table}[h]
\centering
\resizebox{\linewidth}{!}{%
\begin{tabular}{|c|l|}
\hline
\textbf{Term} & \textbf{Description}(in terms of computation time) \\ \hline
\hline
$T_{Enc}$  & Symmetric key encryption/decryption \\ \hline
$T_{SHA}$ & Hash function operations (SHA256) \\ \hline
$T_{ECDH}$ & Elliptic Curve Diffie-Hellman key agreement \\ \hline
$T_{VER}$ & Public key digital signature verification\\ \hline
$T_{ABE-Setup}$ & Key Policy-Attribute Based Encryption Setup\\ \hline
$T_{ABE-KeyGen}$ & Key Policy-Attribute Based Encryption KeyGen\\ \hline
$T_{ABE-Enc}$ & Key Policy-Attribute Based Encryption encryption\\ \hline
$T_{ABE-Dec}$ & Key Policy-Attribute Based Encryption decryption\\ \hline
$T_{IBBE}$ & Identity-Based Broadcast encryption/decryption \\ \hline
$T_{VER}$ & Public key digital signature verification\\ \hline
\end{tabular}
}
\caption{Notation definitions for overhead analysis}
\label{tab:comm-terms}
\end{table}

\begin{table*}[h]
\centering
\resizebox{\linewidth}{!}{%
\begin{tabular}{|c|c|c|c|c|}
\hline
\textbf{Phase} & \textbf{Sensor ($W_i$)} & \textbf{Sensors coordinator ($WNC$)}  & \textbf{Cloud Provider}  & \textbf{User} \\ \hline
\hline
Initialization &  $ T_{ECDH} + T_{Enc}+ n* T_{SHA}$ & $2T_{ECDH} + 2T_{Enc}+ T_{ABE-Setup}+ T_{IBBE}$\ & $T_{ECDH} +T_{Enc}$ & 0 \\ \hline
User registration  & 0 & 0 & $T_{ABE-KeyGen}+T_{VER}$ & $T_{VER}+T_{IBBE}+T_{ABE-KeyGen}$\\ \hline
Data Uploading  & $T_{Enc}$ & $T_{ABE-Enc}+T_{Enc}+2T_{SHA}$ & 0 & 0 \\ \hline
Data Downloading  & 0 & 0 & $T_{VER}$ &  $T_{ABE-Dec}+ T_{Enc}$ \\ \hline
\end{tabular}
}
\caption{Overhead per protocol phase and entity}
\label{tab:comm-cost}
\end{table*}

\subsection{Security Analysis}
In this section, we conduct a comprehensive security analysis of the proposed protocol.

\subsubsection{Mutual Authentication}
A standard Elliptic Curve Cryptography used for authentication and key agreement, with security relying on the hardness of the Elliptic Curve Discrete Logarithm Problem for mutual authentication.

\subsubsection{Secrecy of Key Agreement}
The key agreement handshake between Wis and WNC, as performed in step \ref{key-agg}, employs AES (Advanced Encryption Standard) techniques to share a base value for the key hash chain and other keys generated through a hash algorithm such as SHA-1. The secrecy of the key is contingent upon the difficulty of pre-image computation over a standard hash function.

\subsubsection{Data Confidentiality and Integrity}
Data in our protocol is encrypted with AES, and the symmetric key is encrypted via KP-ABE, relying on the security of AES and KP-ABE for data confidentiality. Message integrity is guaranteed by HMAC values in all data exchanges.

\subsubsection{Fine-Grained Access Control}
Data encryption attributes are established using IBBE and KP-ABE. User roles, set via IBBE, are permanent, while data owners (WNC) can enforce dynamic attributes using KP-ABE. This setup enables precise access control, permitting only users with essential attributes to access specific data.

\subsubsection{Zero Trust Access Control}
As per \cite{blusky}, our protocol integrates criterion-based access control with a multi-dimensional score-based approach for assessing entity trustworthiness. Sensor trust scores are securely stored by WNC using a Collision-Free hash function within a Merkle Tree, preventing sensors from altering their scores with received tokens. WNC reduces sensor scores in response to unauthorized messages, inactivity, or user-reported concerns. The trustworthiness of both the cloud and users is monitored through the application of the UCON+ scheme \cite{tlee}.

\subsubsection{Privacy and Anonymity}
The IBBE scheme is employed to create confidential private keys for system users according to their roles, with user attributes requiring logical conjunction ('AND') for data access. This structure enhances privacy and security by effectively concealing user identities from the public cloud service provider.
\section{Related Work}\label{sec:related}
In \cite{unlesh-iot}, best practices for addressing IoT security challenges are presented. In \cite{blusky}, a Zero Trust framework for IoT access control is introduced. We leverage these frameworks to review related work, with a specific focus on IoT AC \cite{farhadiotac} and Zero Trust AC schemes, given the absence of lightweight real-time Zero Trust AC models, like our proposed scheme.

In \cite{zt-access}, a zero-trust access control system combines ABE with zero-trust scoring for IoT in power-intensive settings. However, it relies on a static, resource-intensive AC module, making it unsuitable for sensor networks. In \cite{zt-context}, a context-aware zero-trust access control framework is designed for healthcare IoT devices. It involves an initial and decision making stages, the former checks basic trust score and bond among resources while the latter powered by Cloud AI analysis. Yet, it's impractical for sensor networks due to non-real-time operation and high resource demands.

In \cite{gope-ac}, a lightweight mutual authentication protocol is proposed for wireless sensor networks but lacks zero trust and robust access control. In \cite{real-time}, an attribute-based access control method is presented for industrial IoT. While it safeguards real-time data integrity, it doesn't adhere to zero trust principles or address device mobility. Table \ref{tab:comparison} summarizes the related work in IoT AC.

Research has focused on secure access control for untrusted storage \cite{cps-survey, ps-lte, 20iot} and the challenges of integrating IoT with cloud computing \cite{iot-cloud-sec, iot-sec-chl, sevr+}. In \cite{blockchain-iot}, blockchain-based models have been proposed; however, their technical characteristics limit their suitability for real-time, resource-limited, dynamic networks.

In the context of fine-grained access control, \cite{hierarchy} introduces a hybrid scheme that combines hierarchical identity-based encryption and CP-ABE for secure cloud data sharing. This approach offers efficient user revocation but raises concerns about potential access by the cloud storage to all data.

\begin{table}[h]
\centering
\resizebox{\columnwidth}{!}{
\begin{tabular}{|c|c|c|c|c|c|c|}
\hline
\textbf{Scheme}  & \textbf{ZT AC} & \textbf{RT AC} & \textbf{Auth} & \textbf{FG AC} & \textbf{Mob.} & \textbf{Light.} \\ \hline
\hline
\textbf{Proposed scheme} & $\checkmark$ & $\checkmark$ & $\checkmark$ & $\checkmark$ & $\checkmark$ & $\checkmark$ \\ \hline
ZT-Access \cite{zt-access} & $\checkmark$ & $\times$ & $\checkmark$ & $\checkmark$ & $\times$ & $\times$ \\ \hline
Khalid et al. \cite{zt-context} & $\checkmark$ & $\times$ & $\checkmark$ & $\times$ & $\times$ & $\times$ \\ \hline
Gope et al. \cite{gope-ac} & $\times$ & $\checkmark$ & $\checkmark$ & $\times$ & $\checkmark$ & $\checkmark$ \\ \hline
Wang et al. \cite{real-time} & $\times$ & $\checkmark$ & $\checkmark$ & $\times$ & $\times$ & $\checkmark$ \\ \hline
\end{tabular}
}
\caption{Comparative Analysis of Access Control (AC) Protocols: Zero Trust AC, Real-Time AC, Authentication, Fine-Grained AC, Sensor Mobility, and Lightweight Computation Support.}
\label{tab:comparison}
\end{table}
\section{Conclusion}\label{sec:conclusion}
Our proposed Zero-Trust real-time lightweight access control protocol offers a pioneering solution to the dynamic and security challenges faced by IoT Sensor Networks. With fine-grained access control, efficient cryptographic design, and a resilient Zero-Trust Access paradigm, it aligns with the stringent requirements of such networks while ensuring data availability and confidentiality. The protocol's performance and security evaluations provide concrete evidence of its practicality and significance, marking a step forward in fortifying the security landscape of IoT Sensor networks.

Future work entails exploring sensor network-specific authentication models \cite{proof-car}, investigating lightweight user attribute revocation mechanisms \cite{kp-revoke}, and implementing the proposed scheme on comparable architectures such as vehicular networks.

\label{lastpage}
\end{sloppypar}


\small
\setlength{\parskip}{-1pt}
\setlength{\itemsep}{-1pt}
\bibliography{paper}
\bibliographystyle{acm}

{
}

\section{Appendix: Cryptographic Preliminaries}\label{sec:appendix}
In this section, we provide an overview of the cryptographic schemes employed in the proposed protocol.

\subsection{Key Policy Attribute-Based Encryption}
KP-ABE is a core public-key cryptography method that allows data owners to encrypt their data, ensuring that only users with matching attributes can decrypt it. It involves four main algorithms: Setup, KeyGen, Enc. and Decryption.

$Setup(k, U)$: This algorithm takes a security parameter (k) and an attribute set (U) as input and generates two sets: the Public Key set (PK) and the Master Key set (MK). The PK is publicly distributed among all parties, while the MK is held in confidence by the authorized entity.

$KeyGen(P, MK)$: This algorithm utilizes the MK and a user's access structure (P) to generate the user's decryption key ($D$).

$Enc(M, I, PK)$: Given the PK, a set of attributes (I), and a message (M), this algorithm produces the ciphertext (E).

$Dec(E, D)$: This algorithm takes the ciphertext (E), and user's decryption key ($D$) as input. If the attribute set (I) aligns with the user's access structure (P), the algorithm decrypts the ciphertext, resulting in the message (M).

\subsection{Identity-Based Broadcast Encryption}
Identity-Based Broadcast Encryption (IBBE) is a cryptographic method for sending encrypted messages over a broadcast channel. Only designated receivers with matching identities can decrypt the data. An IBBE scheme consists of four key algorithms: Setup, KeyExt, Enc, and Decryption.

$Setup(\lambda, l, m)$: This algorithm takes as input the maximum number of receivers (m), the length of each identity string (l), and the security parameter ($\lambda$). It then produces a master key (MK) and a public key set (PK).

$KeyExt(PK, MK, Id_j)$: Given the PK, MK, and a user's identity ($Id_j$), this algorithm generates the user's private key ($SK_{Idj}$).

$Enc(S, PK)$: This encryption algorithm accepts a receiver set (S) and the public key set (PK) as input and returns a header ($hdr$) along with a shared key ($K$).

$Dec(S, Id, SK_{Idj}, hdr, PK)$: The decryption algorithm takes the receiver set (S), a user's identity (id) from the set, their private key ($SK_{Idj}$), the header ($hdr$), and the public key set (PK) as input. It then outputs the shared key ($K$) if the user is authorized to decrypt the message based on their identity.

\end{document}